\documentclass[9pt]{src/IEEEtran}
\IEEEoverridecommandlockouts

\usepackage{cite}
\usepackage{amsmath,amssymb,amsfonts}
\usepackage{algorithmic}
\usepackage{graphicx}
\usepackage{textcomp}
\usepackage{xcolor}
\usepackage{booktabs}
\usepackage{subcaption}
\def\BibTeX{{\rm B\kern-.05em{\sc i\kern-.025em b}\kern-.08em
    T\kern-.1667em\lower.7ex\hbox{E}\kern-.125emX}}
\begin{document}

\title{Bridging Paintings and Music – Exploring Emotion based Music Generation through Paintings

}

\author{
Tanisha Hisariya \qquad Huan Zhang \qquad  Jinhua Liang \\
\textit{ Queen Mary University of London}, Centre for Digital Music, London, United Kingdom \\
t.hisariya@se23.qmul.ac.uk, huan.zhang@qmul.ac.uk, jinhua.liang@qmul.ac.uk
}

\maketitle

\begin{abstract}
Rapid advancements in artificial intelligence have significantly enhanced generative tasks involving music and images, employing both unimodal and multimodal approaches. This research develops a model capable of generating music that resonates with the emotions depicted in visual arts, integrating emotion labeling, image captioning, and language models to transform visual inputs into musical compositions. Addressing the scarcity of aligned art and music data, we curated the Emotion Painting Music Dataset, pairing paintings with corresponding music for effective training and evaluation. Our dual-stage framework converts images to text descriptions of emotional content and then transforms these descriptions into music, facilitating efficient learning with minimal data. Performance is evaluated using metrics such as Fréchet Audio Distance (FAD), Total Harmonic Distortion (THD), Inception Score (IS), and KL divergence, with audio-emotion text similarity confirmed by the pre-trained CLAP model to demonstrate high alignment between generated music and text. This synthesis tool bridges visual art and music, enhancing accessibility for the visually impaired and opening avenues in educational and therapeutic applications by providing enriched multi-sensory experiences.

\end{abstract}


\section{Introduction} \label{sec:intro}
"Art is not what you see but what you make others see." - Edgar Degas. Visual art communicates information and emotions from the artist to the observer, encapsulating cultural and innovative influences from different eras. Similarly, music as an art form evokes a broad spectrum of emotions through its composition and style, paralleling the expressive power of visual arts. This paper explores the innovative intersection of these two art forms by generating music that reflects the emotions perceived in visual artworks such as paintings. This approach not only aims to make art more accessible to the visually impaired by translating visual cues into auditory signals but also extends the research frontier in AI-driven generative models conditioned on images.

Advancements in AI, particularly in generative models, have shown remarkable achievements in mimicking human creativity and generating content that aligns with user expectations across various media, leveraging deep learning techniques~\cite{briot2021artificial,liu2023wavjourney,liang2024wavcraft}. The models capture complex patterns from large datasets and generate outputs that often surpass traditional methods. Music generation is one significant application of AI, utilizing methods to create compositions previously thought unattainable by machines, with techniques categorized into symbolic representation and waveform generation. Symbolic generation focuses on note sequences and events, mainly used by musicians~\cite{vinet2003representation,zhang2024dexter}, while waveform generation produces continuous audio signals interpretable by a general audience, offering a broader application for everyday interactions~\cite{liu2023audioldm,yuan2023leveraging}. Waveform generation's effectiveness heavily relies on the sampling rate to fully capture the audio structure, requiring models that can train effectively on high-dimensional datasets~\cite{colarusso2017raman}.

The challenge of cross-modal generative AI, particularly converting images to music, lies in identifying relationships between these diverse modalities and the scarcity of paired data necessary for training. Despite these challenges, using transfer learning to apply a pre-trained music generation model fine-tuned to specific needs reduces the reliance on large datasets, allowing for the integration of different architectures and modalities to create efficient and seamless hybrid models. This paper aims to bridge the gap between visual and auditory arts, enhancing accessibility and merging distinct forms of artistic expression. The contribution of this work is summarized as follows:
\begin{enumerate}
    \item We proposed a visual-guided music synthesis system that generates music by interpreting the emotions conveyed by images.
    \item Our framework is decomposed into image-to-text and text-to-music tasks, facilitating efficient learning with minimal data. We further enhance training efficiency by exclusively training the decoder in the latent space.
    \item We explore the influence of text descriptions by employing diverse textual conditions. To this end, we have curated for both training and evaluation purposes the Emotion Painting Music Dataset, which is publicly available at this URL~\footnote{https://zenodo.org/records/13717256}.
\end{enumerate}

Our generated music is qualitatively
measured across various metrics using the Fréchet Audio
Distance (FAD)~\cite{kilgour2018fr}, Total Harmonic Distortion (THD), Inception Score
(IS), and KL divergence. Audio-emotion text similarity has also been measured by pre-trained CLAP~\cite{elizalde2023clap} model to demonstrate high alignment between generated music and text. This tool, bridging art and music, holds promise for enhancing learning experiences in educational environments or therapeutic contexts, providing unique multi-sensory engagements.

\section{Related work} \label{sec:related_work}
This research covers three key areas in generative AI: image feature extraction and conversion, deep learning techniques for music generation, and multi-modality in audio and music generation.

\vspace{-0.2cm}\subsection{
Image feature extraction and conversion} \label{subsec:image_feature}
Feature extraction from images can utilize either unimodal or multimodal approaches. The unimodal approach offers straightforward single-label classifications, while multimodal strategies employ language models like LLMs to process visual representations~\cite{li2022blip,liang2023adapting}. Deep neural networks (DNN)s, such as~\cite{he2016deep}, laid the groundwork for the unimodal strategy by categorizing an image to a fixed set of labels. Subsequently, DNNs have been proven to yield the promising performance in other fields, such audio and graphs~\cite{liang2024mind,ding2023acoustic,Zhang2023SymbolicEvaluation}.


The evolution of image captioning leveraged these CNN architectures alongside natural language processing, significantly advanced by the introduction of contrastive learning~\cite{radford2021learning,liang2023adapting}.Contrastive Language-Image Pretraining (CLIP) enhanced the alignment between textual and visual data using dual encoders to calculate cosine similarities between text and image vectors. Following CLIP, BLIP~\cite{li2022blip} further refined the model by integrating a flexible multimodal encoder-decoder framework that excelled in both understanding and generating visual-language tasks with high accuracy. The subsequent introduction of Multimodal large language model~\cite{achiam2023gpt,liang2023acoustic,Liu_2024_CVPR} marked a major advancement, optimizing the handling of more complex contextual and visual information, setting new standards for image captioning capabilities.

\vspace{-0.2cm}\subsection{Deep Learning methods for Music Generation} \label{subsec:music_gen}
The use of Recurrent Neural Networks (RNNs) to generate musical melodies began with Todd~\cite{todd1989connectionist}, but their limited memory capacity led to the development of Long Short-Term Memory (LSTM) units~\cite{hochreiter1997long}, which improved the ability to retain musical sequences over time. The evolution continued with models incorporating Restricted Boltzmann Machines (RBM) for better melody generation~\cite{VanHerwaarden2014PredictingNetworks}, although challenges in long-term memory persisted until Google introduced advancements in RNNs for music with enhanced dependency handling in 2016.

The introduction of generative architectures such as Convolutional Neural Networks (CNNs), Variational Auto Encoders (VAEs), and Generative Adversarial Networks (GANs) marked significant progress. MusicVAE~\cite{roberts2018hierarchical} and MidiNet~\cite{yang2017midinet}, a CNN-based GAN, expanded capabilities for creating coherent musical sequences, although GANs sometimes suffered from mode collapse. MuseGAN~\cite{dong2018musegan} further advanced multi-track music generation, improving the coherence of generated compositions. The advent of WaveNet by Google in 2016 introduced a CNN-based model that employed an autoregressive approach for dynamic raw audio generation, heavily used in both text-to-speech and music generation tasks~\cite{van2016wavenet}.

Significant advances were also seen with the introduction of Music Transformer~\cite{Huang2018MusicTransformer}, which utilized transformers and relative attention models to enhance long-term music generation, surpassing earlier RNN models in efficiency but encountering issues with note redundancy. MuseNet in 2019~\cite{payne2019musenet} and later developments like seqGAN~\cite{yu2017seqgan} and models by Jacek and Teodara using Conditional VAEs and RNNs for emotion-driven music generation~\cite{Mycka2023} highlighted the ongoing challenges of computational demands and complex architecture interpretability while pushing the boundaries of music generation technology.

\vspace{-0.2cm}\subsection{Multi-Modality Audio and Music generation} \label{subsec:multi_modality_music_gen}
In 2022, cMelGAN~\cite{qian2022cmelgan}, a conditional GAN model based on Mel Spectrogram, was introduced to improve music generation efficiency, though it faced challenges with GAN training instability. JukePix~\cite{wang2018jukepix}, a model for converting paintings to music using Convolutional GANs, showed potential in generating multi-track music but was limited by solo performance evaluations. Chen et al. developed MusicLDM~\cite{chen2024musicldm}, a complex model integrating CLAP, VAE, Hifi-GAN, and diffusion models, excelling in music generation but constrained by data sampling rates and computational resources.

Further advancements include AudioLDM~\cite{liu2023audioldm}, which uses CLAP embeddings and a latent diffusion model to achieve high-quality audio generation. Building on this, AudioLDM 2~\cite{Liu2023AudioLDMModels} leveraged GPT-2 to handle various input modalities, showing significant improvements in accuracy. Mousai~\cite{schneider2023mo} and MusicLM~\cite{agostinelli2023musiclm} further explored text-conditioned music generation, with MusicLM providing consistent output despite challenges in processing complex text structures.

MeLoDy by Lam et al.~\cite{lam2024efficient}, used a dual-path diffusion model combined with a language model to enhance semantic modeling and music generation. Despite its innovative approach, training data biases limited its diversity. Sheffer and Adi's im2wav model~\cite{sheffer2023hear}, based on a transformer architecture and CLIP, aimed to generate high-fidelity audio from image inputs but faced computational inefficiencies. Recently, Chowdhury et al. introduced MelFusion~\cite{chowdhury2024melfusion}, synthesizing music from images and text via advanced deep-learning diffusion models, setting new benchmarks in performance and opening avenues for further research in cross-modal generative AI.

\section{Method} \label{sec:method}
The method of generating music from images based on
emotions consists of integrating deep learning models.
Firstly, we will convert the images into their textual
format, and then text along with its associated music will be
used to fine-tune the MusicGen~\cite{Copet2023} model to
generate the required music. By doing this, we are not only
generating the music conditioned on emotions from images,
but we are also exploring the effect of various textual
descriptions during musical generation. In this research
paper, we are exploring four models: an emotion labeling 
model, an image description model, a large language model,
and a Music Generation model. The overview of our
methodology can be seen in Figure 2.

\vspace{-0.2cm}\subsection{Image Emotion Labelling Model} \label{subsec:emotion_labelling}
To effectively perceive the emotions of images, we have
introduced a classification model to label the emotions. The
model will play an important part in determining the
emotions during inference, helping to improve the
consistency and relevance of generated music. Pre-trained
ImageNet ResNet50~\cite{he2016deep} has been chosen as the best model
for this approach due to its ability to manage diverse and
complex datasets with dense layers. Leveraging the technique
of transfer learning, the model has been adapted to a given
dataset with the additional two GRU layers along with a
multi-head Attention layer before the fully connected layer.
The output layer of ResNet50 architecture has been flattening
out to meet the output classes of the dataset. Further, to enhance the performance and to prevent overfitting, some
dropping out of non-essential neurons have been
incorporated before fully connected layers.






\begin{figure}
    \centering
    \includegraphics[width=\linewidth]{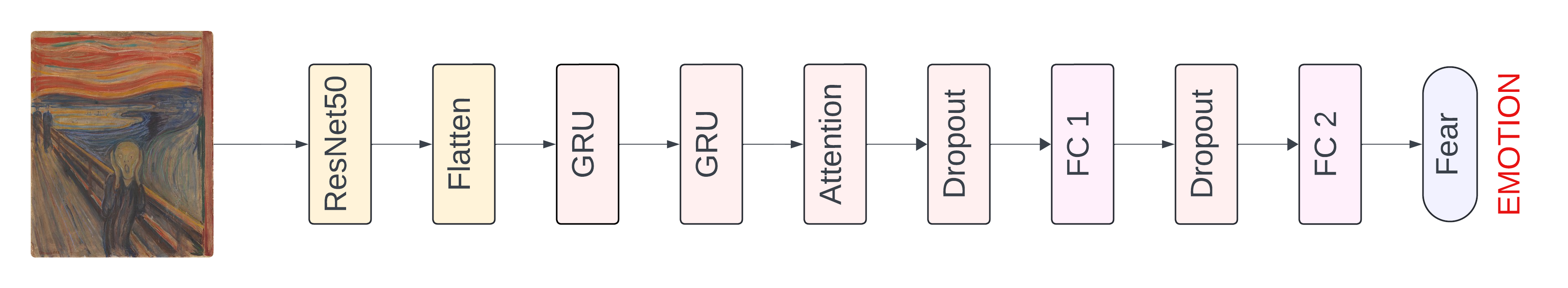}
    \caption{Architecture of Emotion Labelling model with pretrained ResNet50  and additional two bidirectional GRU and one Attention layer proceeding with dropout along fully connected layer.}
    \label{fig:2}
    \vspace{-0.4cm}
\end{figure}

\vspace{-0.2cm}\subsection{Image Description Model} \label{subsec:image_description}
The Image Captioning Model is very crucial as it is responsible for generating the captions of images reflecting emotions perceived by them. By doing this, we aimed to enhance the description of images by extracting more word tokens. We employed BLIP [11], a current state-of-the-art model for Image Captioning, due to its superior performance in generating diverse and descriptive texts aligning closely with visual information.

The model is being conditioned on the emotion labels obtained from the emotion classifier enabling it to give better relevant emotional descriptions. The model is trained on a large amount of highly diversified data so we can directly incorporate the BLIP Large Captioning model to generate the caption. 


\begin{figure*}
    \centering
    \includegraphics[width=0.95\linewidth]{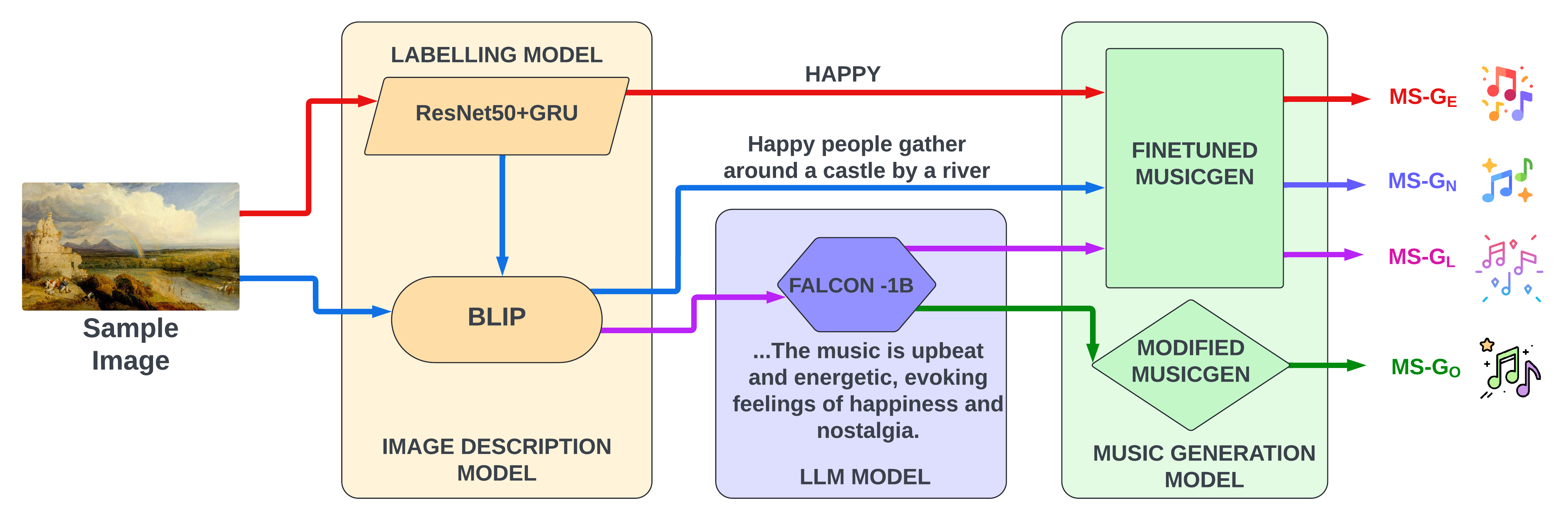}
    \caption{Overview architecture of our model , which encompasses working flow of all the four models represented as different colour giving the output as MS-G$_{E}$ (single label text from ResNet50+ MusicGen), MS-G$_{N}$ (image descritive text from BLIP+MusicGen), MS-G$_{L}$ (Enhanced description from Falcon+ MusicGen), MS-G$_{O}$ (Enhanced description from Falcon+enhanced finetuning method of MusicGen).}
    \label{fig:3}
    \vspace{-0.2cm}
\end{figure*}

\vspace{-0.2cm}\subsection{LLM model} \label{subsec:llm_model}
This model plays an integral role in the evolution of visual and musical information. It further enhances the description generated by the captioning model by incorporating some musical terms that reflect the mood, themes, and musical understanding terms that are very useful for generating the music. There is a need to enhance the description because the Image Captioning Model gives us descriptions based on image features such as objects, colors, and more, while the models for generating music need some music related component details in that to optimally perform. Providing this type of description leads to a better quality and resemblance of music. That’s why we are integrating the LLM model into our framework to fill the gap between the provided description and the expected inputs. The model not only works with the optimization of description but also ensures that the information about the visual image, especially the essence of emotion, is not lost. 

The LLM model we have incorporated here is FalconRW-1B \cite{penedo2023the} because it is fully open-source, and able to perform well even with restricted computational power. It consists of language modelling decoder-based architecture that is only incorporated with many advanced techniques like Attention.  The input requirement of this model is characterized into three parts as shown in Table 1: system message (intent to tell the behavior of the model), instructions (intent to give the proper input), and response (the response model is providing).


\begin{table}[ht]
\centering
\caption{Description of input format given to falcon 1B model, emphasizing the system message and instruction, followed by a description provided.}
\label{tab:description}
\begin{tabular}{@{}p{2cm}p{6.5cm}@{}}
\toprule
Role & Content \\ 
\midrule
System Message & 
  You are an enhanced description generator. You will be given with image description and you have to enhance those in musical terms. \\

Instruction & 
  Generate a musical theme description for the following image description: "\textit{sad man in a sailor’s hat sitting at a table}". Include details like mood, genre, tempo, and melody in 2 lines. \\ 
\bottomrule
\end{tabular}
\vspace{-0.25cm}
\end{table}

\vspace{-0.2cm}\subsection{ Music Generation } \label{subsec:music_gen}
The MusicGen-small model, abbreviated as MG-S, was fine-tuned for generating music based on various textual inputs derived from corresponding image-to-text models and audio files. Text and audio files were encoded into tokens using specialized encoder models, followed by a conditioning and fusion process involving an Attention mechanism. This was further processed by a transformer model using masking techniques to generate relevant tokens for loss computation and parameter refinement.

Our experimental process involved several iterations of the MG-S model to enhance music generation capabilities:

\textbf{MG-S Emotive}: Utilized emotion descriptions from the Image Emotion Labeling model to adjust the model's parameters.

\textbf{MG-S Narrative}: Conditioned the music generation on image descriptions enriched with emotional cues, with comprehensive parameter tuning in the model.

\textbf{MG-S Lyrical}: Enhanced the descriptive content of images using a Language Model that incorporates musical knowledge, thoroughly fine-tuning the model.

\textbf{MG-S Optimized}: Introduced architectural improvements to optimize model performance. These improvements included pre-processing music and text files prior to training to stabilize and streamline the training process. The approach involved storing precomputed tensors, freezing initial layers to enhance stability, and modifying input handling for consistent learning. This version also adjusted how input descriptions were managed to ensure accurate performance evaluation on our dataset.

These versions of MG-S represent a progression in the methodical enhancement of music generation, conditioned on textual and emotional cues, leading to a refined and efficient training methodology.

\section{Experiments} \label{sec:experiments}
\vspace{-0.2cm}\subsection{Datasets} \label{subsec:datasets}
Data collection and preparation are elementary steps in developing any model. Due to the lack of an existing paired dataset encompassing both art and music, which share the same emotional attribute, we proposed to make our own bespoke paired dataset integrating two different art forms, painting and music, while conveying the same emotion. 

For the image paintings dataset, we utilized WIKIART EMOTION DATASET~\cite{mohammad2018wikiart}, an openly available dataset of wiki art depicting various emotions. Wikiart is a collection of various paintings that evolved from different eras of the past to present, symbolizing different art forms and meanings. These paintings have been analyzed and further categorized into more than ten emotions in the emotion dataset. Based on these, we have manually analyzed datasets for five particular emotions - Happy, Angry, Sad, Fun, and Neutral and collected 1200 such paintings, which will serve as one part of our paired dataset. The manual selection process depicts the accurate representation of each dataset conveying the emotion.

The next part contains the collection of music datasets that should convey the same emotion. For this purpose, we selected MIREX EMOTION DATASET, a dataset of 193 MIDI files of music depicting various emotions. Initially, we preprocess the musical dataset from its raw form of  MIDI and convert it into .wav form at 32KHz making them compatible frequency for the MusicGen model. We further combined the emotional aspects of several parts to categorize it into five similar emotions as with paintings finally. Furthermore, as we are using a fixed 30s audio segment chunk in our music generation model, the music has been trimmed into various 30s without any overlapping between two different audios. By doing these preprocessing steps, we can generate more music samples that are uniquely identified from each other. 

Later, to form the final paired dataset, the music and paintings have been associated with each other randomly, depicting the same emotions. In this way, we can have 1200 different pairs of paintings and music depicting five different emotions. Furthermore, for the training purpose of the model, we took around 80\% of data from each section of emotion, with the remaining 20\% split evenly between evaluation and testing sets.

\vspace{-0.2cm}\subsection{Evaluation metrics} \label{subsec:metrics}
To evaluate the quality and resemblance of generated music, we used a set of objective evaluation metrics to measure quality, smoothness, noise, and distortion in the generated music:

\textbf{Frechet Audio Distance (FAD)}.
FAD, inspired by the Frechet Inception Distance, measures the similarity between the statistical distributions of generated and reference music sets using the VGGish model for feature extraction. A lower FAD score indicates greater similarity to the reference set.

\textbf{Contrastive Language Audio Pretraining (CLAP)}.
CLAP score calculates the similarity between text descriptions and audio using embeddings from the pre-trained LAION CLAP model's text and audio encoders, computing cosine similarity between them. A higher CLAP score indicates better alignment between text and audio.

\textbf{Total Harmonic Distortion (THD) score}.
THD measures the harmonic distortion present in the generated music, assessing the purity of the audio signal. Lower THD scores signify less distortion and higher audio quality.

\textbf{Inception score (ISc)}.
IS reflects the variety and diversity of the generated audio group. A higher Isc indicates a diverser distribution of synthesise music. 

\textbf{Kullback-Leibler (KL) divergence}.
KL divergence quantifies the difference between the probability distributions of reference and generated features. A lower KL score indicates closer resemblance between two distributions, suggesting better generation fidelity.

\vspace{-0.2cm}\subsection{Training} \label{subsec:training}
All our experiments were performed on one NVIDIA A40 GPU with 48 GB of memory. This setup allowed us to perform training with a batch size of 16, using the small version of the MusicGen model over 40 epochs. We used an AdamW optimizer with early stopping and a learning rate of 1e-5, a cosine scheduler with warmup steps of 100. During inference, we took the $\text{top}_k$ as 250, selecting the top 250 most resembling audio tokens at a temperature of 1. 


\vspace{-0.2cm}\subsection{Results} \label{subsec:results}
\begin{table}[]
\centering
\caption{Objective comparison of test set data for emotion based image to music generation across all the models. Here the best results are made \textbf{bold}.}
\label{tab:results}
\begin{tabular}{@{}lllllll@{}}
\toprule
Model         & FAD$\downarrow$ & CLAP$\uparrow$ & KL$\downarrow$ & THD$\downarrow$ & ISc$\uparrow $ \\ 
\midrule
MG-S Emotive  & 7.02            & 0.075          & 0.054          & 1.79          & \textbf{1.044} \\
MG-S Narrative & 5.22           & 0.096          & 0.045          & \textbf{1.73}          & 1.032          \\
MG-S Lyrical   & \textbf{5.06}  & 0.11           & 0.046          & 1.92 & 1.031          \\
MG-S Optimized & 5.54           & \textbf{0.13}  & \textbf{0.012} & 1.75          & 1.033          \\
\bottomrule
\end{tabular}
\vspace{-0.2cm}
\end{table}


The architectural experiments with the four variants of the MusicGen model, as reflected in Table~\ref{tab:results}, offer detailed insights into each model's performance improvements and challenges. MG-S Emotive, our baseline model, uses ResNet50 and GRU to extract single-word emotion labels, demonstrating limitations with high FAD and KL scores, and low CLAP scores indicating poor text-to-music alignment and outputs with significant noise. MG-S Narrative advances this by using the BLIP model for richer emotional captioning, improving text-music alignment as seen in the higher CLAP scores and reduced noise, though it struggles with complex emotions like anger. MG-S Lyrical incorporates an LLM to enrich musical context in text descriptions, enhancing semantic appropriateness and improving FAD and CLAP scores but facing challenges in balancing complexity with fidelity as indicated by the KL and THD scores. Finally, MG-S Optimized integrates enriched contextual descriptions with a modified tuning pipeline, significantly reducing training times and achieving the highest CLAP scores while minimizing distortion and noise, demonstrating the most effective architecture in complex emotional contexts as evidenced by the spectrogram analysis. These progressive refinements highlight the enhanced capabilities of MusicGen in generating high-fidelity music aligned with complex textual descriptions.

\begin{figure}
    \centering
    \includegraphics[width=\linewidth]{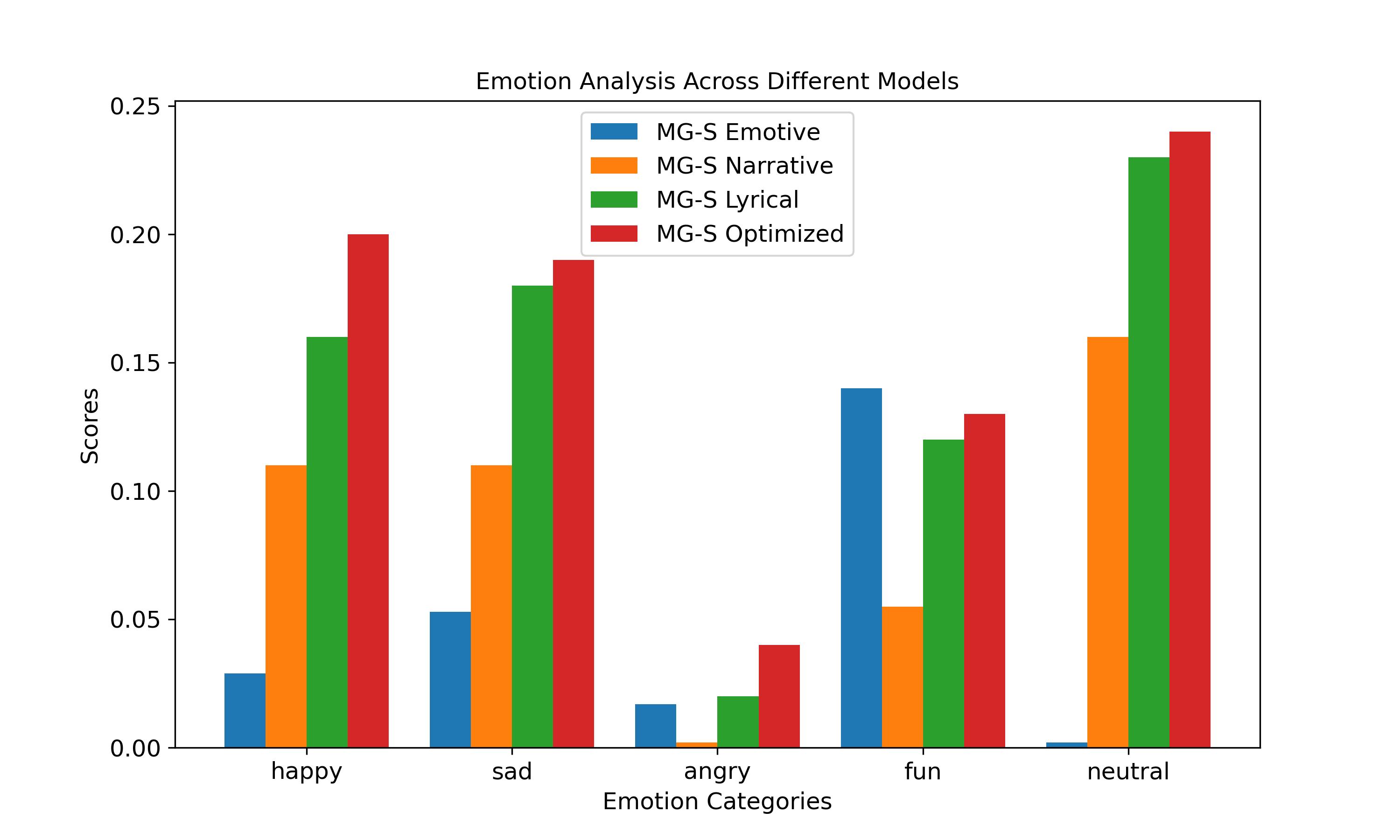}
    \caption{CLAP analysis of generated song across model with their emotions. It is being meausred with providing “{emotion} song” as text  during CLAP calculation.}
    \label{fig:4}
    \vspace{-0.4cm}
\end{figure}

\section{Conclusion} \label{sec:conclusion}
This work introduced a generative model to produce music conditioned on the emotions depicted in paintings, serving as a step towards integrating visual art and music through technology. The model demonstrates the feasibility of using generative AI to create emotionally resonant music, addressing a notable gap in modality conversion. Our evaluation focused on the quality, diversity, and presence of noise in the generated music, highlighting the discrepancies between ideal model inputs and typical user-provided data. Addressing these discrepancies is crucial for improving output quality.

The research also points out the limited availability of datasets appropriate for training art-music generation models and suggests enhancing dataset diversity for better model training. The findings reveal significant gaps in how the models interpret single-label and non-musical descriptions compared to user expectations, underscoring the need for more sophisticated handling of input data. Moreover, the study identifies the high inference time of the model as a challenge for real-time application, suggesting further optimization is needed. Future work should explore developing specific evaluation metrics tailored to this multimodal context to enhance the precision of assessments, potentially advancing the field of text-conditioned generative models.


\newpage
\bibliographystyle{src/IEEEtran}
\bibliography{src/reference}

\end{document}